\documentclass[journal]{IEEEtran}

\ifCLASSINFOpdf
  \usepackage[pdftex]{graphicx}
\else
\fi
\usepackage{booktabs}
\renewcommand{\arraystretch}{1.3}
\usepackage{multirow}
\usepackage{array}
\usepackage{xcolor}
\usepackage{amsmath}
\usepackage{balance}

\newcolumntype{P}[1]{>{\centering\arraybackslash}p{#1}}

\begin{document}

\title{U-FedTomAtt: Ultra-lightweight Federated Learning with Attention for Tomato Disease Recognition}

\author{Romiyal~George,
        Sathiyamohan~Nishankar,
        Selvarajah~Thuseethan,~\IEEEmembership{Member,~IEEE,}
        Chathrie Wimalasooriya,
        Yakub~Sebastian,
        Roshan~G.~Ragel,~\IEEEmembership{Member,~IEEE,}
        and~Zhongwei~Liang,~\IEEEmembership{Member,~IEEE}
\thanks{R. George, S. Nishankar and R. G. Ragel are with the Department of Computer Engineering, University of Peradeniya, Sri Lanka.}
\thanks{S. Thuseethan and Y. Sebastian are with the Faculty of Science and Technology, Charles Darwin University, Australia.}
\thanks{C. Wimalasooriya is with the Canadian Institute for Cybersecurity, University of New Brunswick, Canada.}
\thanks{Z. Liang is with the School of Mechanical and Electrical Engineering, Guangzhou University, China.}

\thanks{Manuscript received Nov 5, 2025; revised MM DD, YYYY.}}

\markboth{Preprint Under Review}%
{Shell \MakeLowercase{\textit{et al.}}: Bare Demo of IEEEtran.cls for IEEE Journals}

\maketitle

\begin{abstract}
Federated learning has emerged as a privacy-preserving and efficient approach for deploying intelligent agricultural solutions. Accurate edge-based diagnosis across geographically dispersed farms is crucial for recognising tomato diseases in sustainable farming. Traditional centralised training aggregates raw data on a central server, leading to communication overhead, privacy risks and latency. Meanwhile, edge devices require lightweight networks to operate effectively within limited resources. In this paper, we propose \textit{U-FedTomAtt}, an ultra-lightweight federated learning framework with attention for tomato disease recognition in resource-constrained and distributed environments. The model comprises only 245.34K parameters and 71.41 MFLOPS. First, we propose an ultra-lightweight neural network with dilated bottleneck (DBNeck) modules and a linear transformer to minimise computational and memory overhead. To mitigate potential accuracy loss, a novel local-global residual attention (\texttt{LoGRA}) module is incorporated. Second, we propose the federated dual adaptive weight aggregation (\texttt{FedDAWA}) algorithm that enhances global model accuracy. Third, our framework is validated using three benchmark datasets for tomato diseases under simulated federated settings. Experimental results show that the proposed method achieves 0.9910\% and 0.9915\% Top-1 accuracy and 0.9923\% and 0.9897\% F1-scores on SLIF-Tomato and PlantVillage tomato datasets, respectively.

\end{abstract}


\IEEEpeerreviewmaketitle

\section{Introduction} \label{sec:introdution}
\IEEEPARstart{F}{ederated} learning (FL) and lightweight neural networks have independently achieved success across diverse domains. FL is widely applied in healthcare \cite{abbas2024federated}, finance \cite{awosika2024transparency} and mobile systems \cite{lee2024federated} for privacy-preserving collaborative modelling. Similarly, lightweight neural networks perform effectively in resource-constrained settings, where computational efficiency is critical \cite{liu2024lightweight}. Unlike other domains, agriculture uniquely supports the combined use of FL and lightweight neural networks. In automated plant disease recognition, data from geographically dispersed farms render centralised aggregation impractical and privacy-sensitive. In fact, this is particularly relevant for tomato, a globally cultivated crop with disease data available from multiple regions \cite{george2025past}. Moreover, deploying plant disease recognition models in rural and edge environments requires lightweight architectures optimised for limited computational and energy resources. The integration of FL with lightweight neural networks, therefore, presents a promising solution for achieving accurate, privacy-preserving and scalable plant disease recognition in agricultural settings.

In the agriculture domain, an ideal plant disease recognition system should (i) collaboratively learn from distributed farms while preserving data privacy and (ii) operate efficiently in terms of computation and memory on resource-constrained edge devices. However, existing plant disease recognition approaches typically focus on either centralised deep learning, which demands extensive data transfer and computation \cite{shoaib2023advanced}, or lightweight models that cannot fully exploit distributed data across farms \cite{wei2025lightweight}. Consequently, existing plant disease recognition approaches exhibit a notable gap in simultaneously addressing data privacy, distributed learning and computational efficiency. The integration of lightweight neural networks with FL has shown promising results in other domains and can be applied to plant disease recognition for globally cultivated crops, such as tomato.

To address these limitations, this study proposes an ultra-lightweight FL framework for tomato leaf disease recognition. In most cases, reducing the architectural complexity of deep neural networks to enhance computational efficiency often leads to a degradation in recognition accuracy \cite{abdusalomov2025optimized}. Several approaches have been investigated to improve the performance of lightweight neural networks. Among these, attention mechanisms have consistently achieved superior results across diverse domains \cite{takalkar2021lgattnet}. In the agricultural domain, the incorporation of attention modules into lightweight architectures has similarly yielded substantial gains in classification accuracy and generalisation performance \cite{duhan2025rtr_lite_mobilenetv2, janarthan2025efficient}. These improvements are primarily characterised by the ability of attention mechanisms to selectively emphasise informative features while suppressing irrelevant background noise, which leads to more discriminative representations.

\begin{figure*}[h]
    \centering
    \includegraphics[width=\linewidth]{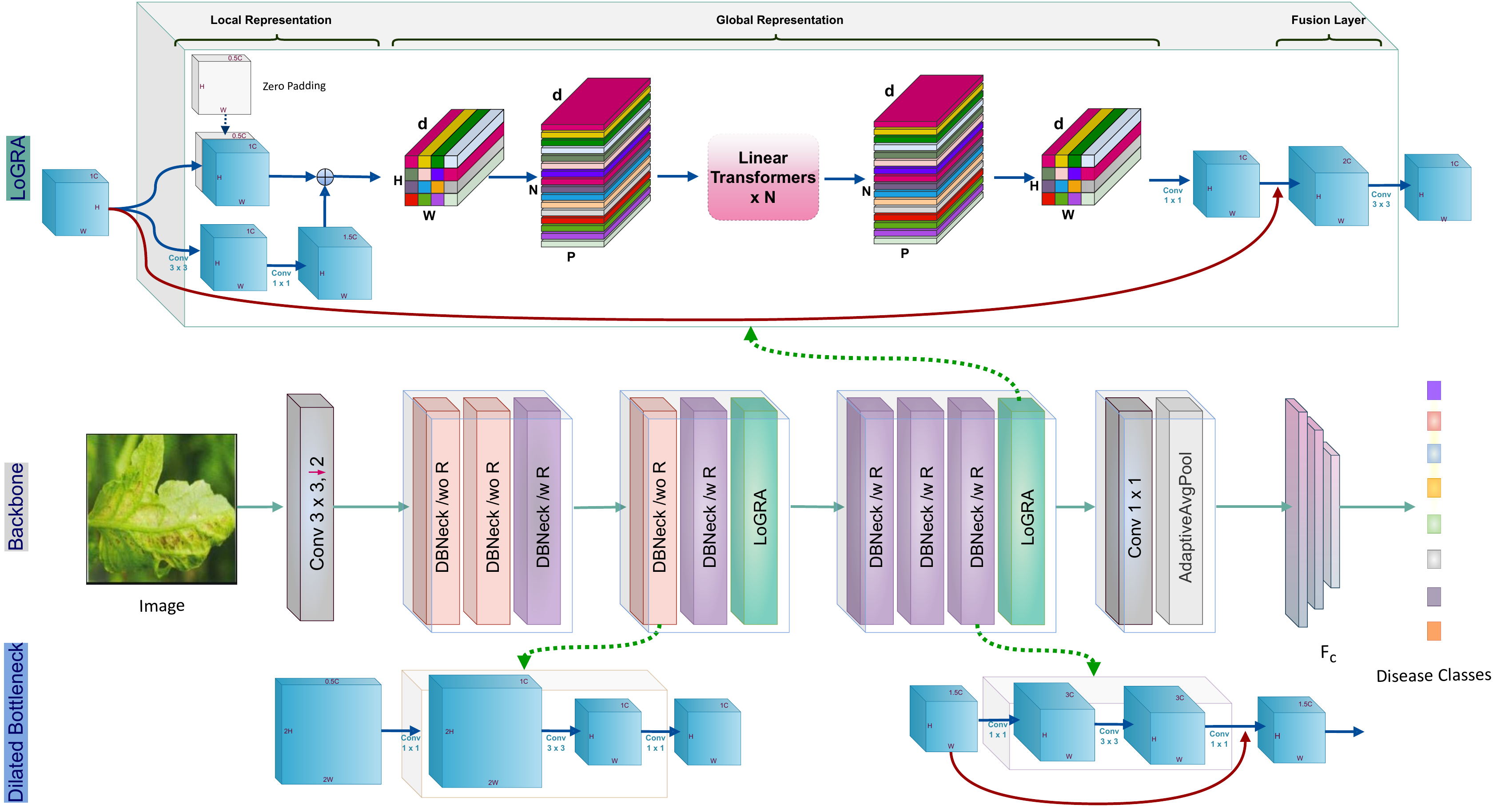}
    \caption{Illustration of the proposed ultra-lightweight neural network with \texttt{LoGRA} for tomato leaf disease recognition. The red arrows denote residual connections, the blue arrows represent standard connections, and the green dashed lines indicate the expansion of the layer architecture.}\label{fig:ufedtomatt}
\end{figure*}

Integrating lightweight architectures within FL enables privacy-preserving and computationally efficient training using distributed data. Nevertheless, deploying such models in FL environments presents significant challenges, such as data heterogeneity, statistical imbalance and variability in client performance. In-field plant disease datasets, particularly those for tomato, inherently exhibit a non-independent and identically distributed (non-iid) structure due to variations in environmental conditions, imaging angles and disease severity. These issues are further compounded by the limitations of conventional aggregation strategies, such as FedAvg \cite{mcmahan2017communication}, FedProx \cite{li2020federated}, FedNova \cite{wang2020tackling}, Scaffold \cite{karimireddy2020scaffold}, MOON \cite{li2021model} and Per-FedAvg \cite{fallah2020personalized}, which are unable to effectively account for both differences in dataset size and local model performance. In fact, none of these approaches effectively identifies the trade-off between dataset size and local model performance, a factor that significantly influences the overall accuracy of the global model.

To address the shortcomings of existing lightweight FL frameworks in managing highly non-iid and distributed data related to tomato leaf disease, we propose a set of targeted enhancements. First, we propose an attention-enhanced ultra-lightweight neural network designed for seamless integration within FL environments. Second, we develop a novel aggregation algorithm that adaptively balances client contributions by accounting for both dataset quality and local model performance. In summary, our study has three main contributions: 

\begin{enumerate}
    \item We propose an ultra-lightweight neural network (shown in Fig. \ref{fig:ufedtomatt}) incorporating a novel attention mechanism, namely \textit{local-global residual attention} (\texttt{LoGRA}), which effectively captures informative features while suppressing irrelevant background noise.

    \item Through the proposed \textit{Federated Dual Adaptive Weight Aggregation} (\texttt{FedDAWA}), as elaborated in Subsection~\ref{sec:feddawa}, we introduce a principled FL aggregation approach that adaptively weights client updates based on each client’s data size and model performance.
    
    \item We develop an ultra-lightweight FL framework, namely \texttt{U-FedTomAtt}, that combines \texttt{LoGRA} and \texttt{FedDAWA} to achieve efficient tomato leaf disease recognition under intensely non-iid and distributed settings.

    \item Extensive experiments on a highly non-iid tomato leaf disease dataset demonstrate that the proposed \texttt{U-FedTomAtt} framework attains higher accuracy, faster convergence and lower computational cost than state-of-the-art baselines.
\end{enumerate}

The remainder of this paper is organised as follows. Section \ref{sec:relatedwork} reviews related studies on lightweight and FL approaches for plant disease recognition. Section \ref{sec:proposedmethod} describes the proposed \texttt{U-FedTomAtt} framework and its main components. Section \ref{sec:experiments} presents the experimental setup, results and performance analysis. Finally, Section \ref{sec:conclusion} concludes the paper and highlights directions for future work.

\section{Related Work} \label{sec:relatedwork}
Lightweight FL has become a key research focus, which is driven by the need for efficient, privacy-preserving and distributed training on resource-limited agricultural devices. Attention-based lightweight models have been widely explored to reduce complexity while maintaining accuracy. Likewise, numerous studies have applied FL in agriculture for decentralised training across farms and devices. This review surveys major works on attention-aware lightweight neural networks (see Subsection \ref{sec:lightweightapproaches}) and FL (see Subsection \ref{sec:FLapproaches}) in tomato disease recognition. 

\subsection{Attention-aware Lightweight Neural Networks} \label{sec:lightweightapproaches}
Several lightweight models have demonstrated strong performance in plant disease recognition. For example, AgriFusionNet, based on EfficientNetV2-B4, integrates multimodal data to achieve 94.3\% accuracy with 28.5 ms inference time, while using 30\% fewer parameters than larger models such as Vision Transformer and InceptionV4 \cite{albahli2025agrifusionnet}. LWDN, a pruned DenseNet121 variant, achieves 99.37\% accuracy with 1.5M parameters, making it 93\% smaller than InceptionV3 and Xception and ideal for mobile deployment \cite{dheeraj2024lwdn}. In the context of tomato leaf disease recognition, a Siamese network-based lightweight framework with approximately 2.96 million trainable parameters has been proposed \cite{thuseethan2024siamese}. In \cite{zhong2023lightmixer}, LightMixer, a lightweight convolutional neural network (CNN) with only 1.5M parameters, achieved 99.3\% accuracy on a publicly available dataset. In \cite{das2025xltldisnet}, a lightweight approach, namely XLTLDisNet, was proposed, which demonstrates an accuracy of 97.24\% on the PlantVillage dataset for tomato leaf disease classification. The majority of lightweight models demonstrate inferior performance compared to deeper neural architectures, particularly under real-world in-field conditions.

Attention has been increasingly recognised as a promising solution to mitigate recognition performance degradation. Integrating channel, spatial, or hybrid attention mechanisms into lightweight neural networks allows models to focus on the most discriminative features of diseased plant images \cite{wang2024attention}. Models such as LBFNet \cite{chen2023lbfnet} and LDAMNet \cite{zhang2024lightweight} have achieved higher accuracy in tomato disease recognition while significantly reducing model size and inference time compared to models without attention integration. These networks employ multi-channel or dual-attention mechanisms to capture fine-grained disease features and suppress background noise for enhanced recognition performance. For instance, the use of the Convolutional Block Attention Module (CBAM) in lightweight CNNs improved average accuracy while achieving a 16-fold reduction in parameters compared to standard ResNet50 \cite{bhujel2022lightweight}. For the first time, in \cite{sun2025ultra}, an ultra-lightweight recognition model with five inverted residual modules and an efficient attention mechanism was developed, effectively balancing model complexity and accuracy. These advancements establish attention-based lightweight networks as a practical and effective solution for real-time tomato leaf disease recognition.

\subsection{FL Approaches} \label{sec:FLapproaches}
FL enhances data privacy while enabling the use of distributed datasets otherwise restricted by privacy or regulatory constraints \cite{ali2025recent}. Specifically, FL is valuable in domains with sensitive or widely distributed data, such as healthcare, finance, and agriculture. In plant disease recognition, FL addresses key challenges, with distributed data across geographically dispersed farms improving recognition accuracy and reducing communication costs \cite{mohammed2025comprehensive}. Traditional centralised approaches face logistical and privacy challenges, whereas FL trains local models on region-specific data to construct a generalised global model. Recent FL frameworks using intelligent weight transferring and hierarchical CNNs achieve up to 97.5\% accuracy, surpassing FedAvg and FedAdam \cite{hari2024improved}. FL also handles non-independent and identically distributed (Non-IID) data, a common challenge in agriculture due to regional variations in crops, disease prevalence, and imaging conditions. Moreover, attention-based lightweight CNNs in FL enhance accuracy while maintaining low communication and computational overhead \cite{hari2025adaptive}.

FL shows considerable potential in tomato leaf disease diagnosis and classification. Studies demonstrate that federated CNN models can accurately identify and categorise tomato leaf diseases across multiple severity levels, achieving high accuracy (96–99\%) and strong precision, recall, and F1-scores, even with data distributed across multiple clients. For instance, in \cite{mehta2023federated}, an FL-based CNN model for tomato leaf disease recognition achieved 96–98\% accuracy. Lamani et al. \cite{lamani2024tomato} proposed a CNN with FL, achieving 98–99\% overall accuracy across all clients and classes. These models generally maintain consistent performance across clients and disease classes, demonstrating robustness and generalisability. More importantly, FL facilitates the development of effective diagnostic tools for tomato leaf disease while preserving farmers’ data privacy and leveraging distributed datasets \cite{trivedi2023tomato}.

The literature shows that lightweight, attention-based CNNs achieve high accuracy with reduced parameters, but their performance often declines under real-world in-field conditions \cite{albahli2025agrifusionnet,dheeraj2024lwdn,thuseethan2024siamese,das2025xltldisnet}. Attention mechanisms improve feature discrimination and suppress background noise, enhancing recognition performance \cite{wang2024attention,chen2023lbfnet,zhang2024lightweight,bhujel2022lightweight,sun2025ultra}. FL preserves data privacy, leverages distributed datasets, handles Non-IID data and maintains high accuracy across multiple clients \cite{ali2025recent, mohammed2025comprehensive, hari2024improved, hari2025adaptive, mehta2023federated, lamani2024tomato, trivedi2023tomato}. However, existing FL models often rely on relatively heavy architectures or lack efficient attention mechanisms, limiting real-time deployment on resource-constrained agricultural devices. Motivated by these gaps, the proposed framework combines an ultra-lightweight attention-based neural network with FL to achieve high accuracy and low computational overhead for tomato leaf disease recognition.

\section{Proposed Method} \label{sec:proposedmethod}
In this section, we present the core components of the proposed \texttt{U-FedTomAtt} framework, which consist of the novel \texttt{FedDAWA} algorithm designed for the FL setting and an ultra-lightweight attention-based neural network backbone used within the overall framework.

\subsection{Problem Setting} \label{sec:problemsetting}
In this work, we consider an FL setting consisting of a central server and $K$ participating clients, denoted as $\mathcal{C} = \{1, 2, \dots, K\}$. Each client $k$ possesses a local dataset 
$\{(x_i^k, y_i^k) \}_{i=1}^{N_k}$ 
with $N_k$ samples, where $x_i^k$ represents the input data and $y_i^k$ denotes the corresponding label. The total number of samples across all clients is 
$N_{total} = \sum_{k=1}^{K} N_k$. The objective of the FL process is to collaboratively learn a shared global model $w$ by minimising the aggregated global loss function:

\begin{equation}
    F(w) = \sum_{k=1}^{K} \alpha_k F_k(w)
\end{equation}
where, \( \alpha_k \) represents the aggregation weight of client \( k \) determined by the proposed \texttt{FedDAWA} algorithm, which is explained in Subsection \ref{sec:feddawa}. The term \( F_k(w) \) denotes the local objective function for client \( k \), which is defined as:
\begin{equation}
    F_k(w) = \frac{1}{N_k} \sum_{i=1}^{N_k} \mathcal{L}(f(x_i^k; w), y_i^k)
\end{equation}
where \( f(x_i^k; w) \) is the model prediction for the input sample \( x_i^k \) with parameters \( w \), \( y_i^k \) is the corresponding ground-truth label, and \( \mathcal{L}(\cdot) \) denotes the local loss function, computed using the cross-entropy criterion.

\subsection{FedDAWA} \label{sec:feddawa}
\begin{figure}[ht]
    \centering
    \includegraphics[width=\linewidth]{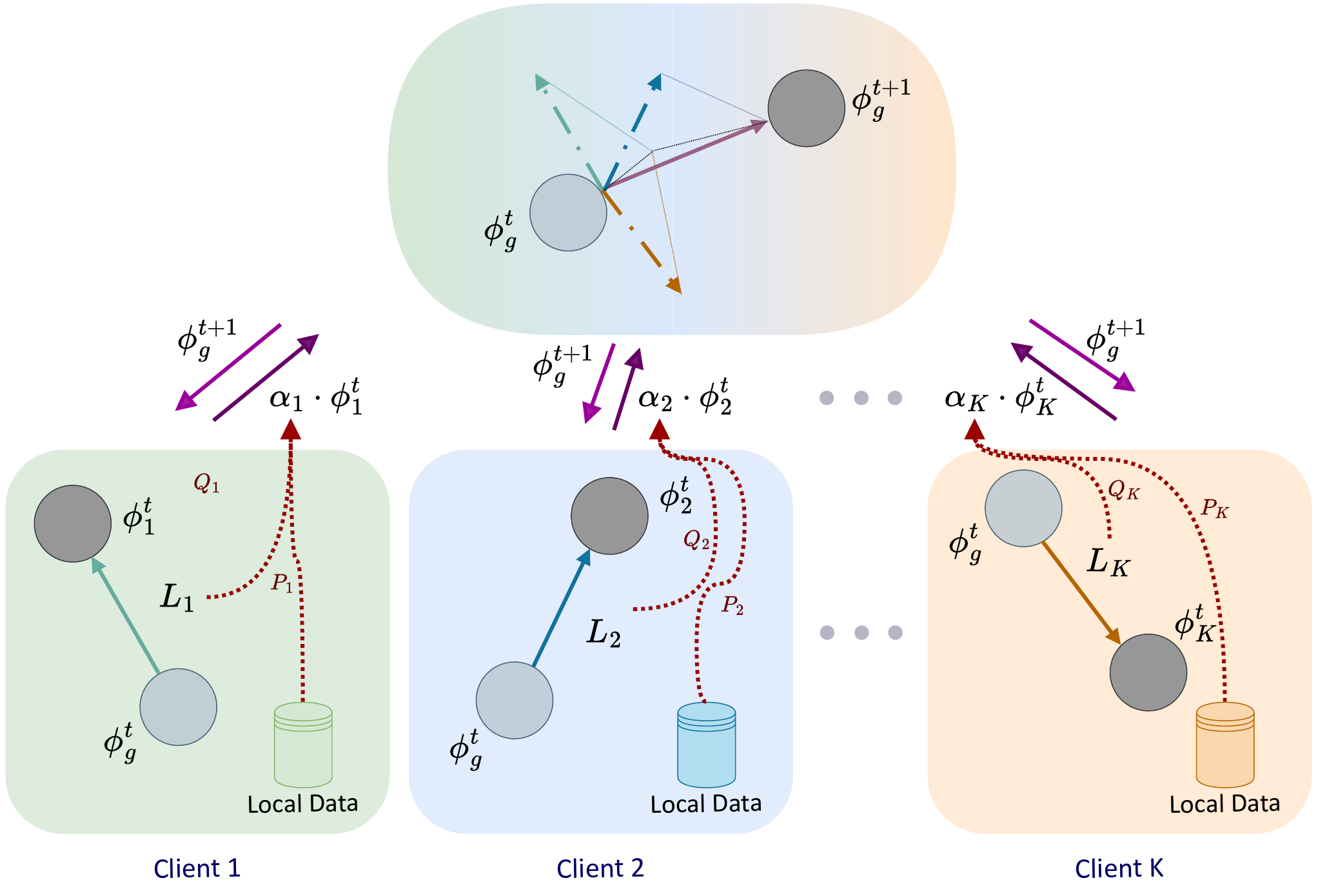}
    \caption{Illustration of \texttt{FedDAWA} algorithm.}\label{fig:feddawa}
\end{figure}

We propose a novel adaptive aggregation mechanism, federated dual adaptive weight aggregation \texttt{FedDAWA}, which jointly considers client dataset size and the inverse of local validation loss to achieve a more balanced, fair and performance-aware global model aggregation. Fig. \ref{fig:feddawa} illustrates the \texttt{FedDAWA} approach for $K$ clients. The principal contribution lies in computing the aggregation weight \( \alpha_k \) for each client $k$, as a convex combination of two components: the \textit{data size component} ($P_k$) and the \textit{inverse validation loss component} ($Q_k$). The \texttt{FedDAWA} algorithm introduces an adaptive weighting coefficient $\beta$ to dynamically regulate the trade-off between these two factors during global aggregation.

In contrast to the conventional FedAvg \cite{mcmahan2017communication} algorithm, which relies solely on the dataset size for client contribution, \texttt{FedDAWA} considers both the quantity of data and the quality of local learning. This design enables clients with larger and better-performing local models to have a proportionally greater impact on the global update, which in turn improves training stability and convergence under non-iid conditions.

\subsubsection{Data Size Component ($P_k$)}
This component captures the relative size of local training data for each client $k$:
\begin{equation}
    P_k = \frac{N_k}{N_{total}}
\end{equation}
with $N_k$ denoting the local dataset size of client $k$.

\subsubsection{Inverse Validation Loss Component ($Q_k$)}
To account for local model quality, the second component utilises the inverse of the local validation loss $L_i$ after $E$ epochs of local training. Clients achieving lower losses are given proportionally higher weights. The component is normalised across all clients as follows:

\begin{equation}
    Q_k = \frac{1 / (L_k + \epsilon)}{\sum_{j=1}^{K} 1 / (L_j + \epsilon)}
\end{equation}
where, $\epsilon = 10^{-6}$ ensures numerical stability.

\subsubsection{Adaptive Aggregation Formulation}
The aggregation weight \( \alpha_k \) assigned to client $k$ is defined as a convex combination of two normalised components:
\begin{equation}
    \alpha_k = \beta \cdot P_k + (1 - \beta) \cdot Q_k
\end{equation}
where, the adaptive weighting coefficient $\beta \in [0, 1]$ regulates the relative contribution of the two components. When $\beta = 1$, the aggregation is entirely determined by the dataset size component. In contrast, when $\beta = 0$, the aggregation depends exclusively on model performance, quantified by the inverse of the local validation loss. For intermediate values $0 < \beta < 1$, the aggregation adaptively balances the influence of both factors, achieving an optimal trade-off between data quantity and learning quality across clients.

\subsubsection{Global Aggregation}
After computing $\alpha_k$ for each client, the global model parameters are updated using the adaptively weighted aggregation as given below:

\begin{equation}
F(w) = \sum_{k=1}^{K} \left( \frac{\beta \cdot P_k + (1 - \beta) \cdot Q_k}{\sum_{j=1}^{K} (\beta \cdot P_j + (1 - \beta) \cdot Q_j)} \right) F_k(w)
\end{equation}

\begin{table*}[ht]
\centering
\caption{Detailed configuration of the proposed ultra-lightweight neural network with LoGRA.}
\label{tab:ultralightweightnetwork}
\setlength{\tabcolsep}{4pt}
\begin{tabular}{|c|l|c|c|c|c|c|}
\hline
\textbf{\#} & \textbf{Layer} & \textbf{Input Shape} & \textbf{[Kernel Size, Stride]} & \textbf{Padding/Dilation} & \textbf{Activation} & \textbf{Output Shape} \\ 
\hline
1 & Conv2D & 256$\times$256 & [3$\times$3, 2] & padding=1 & ReLU & 128$\times$128 \\
2 & DBNeck \textit{/wo R} & 128$\times$128 & [1$\times$1 $\rightarrow$ 3$\times$3 $\rightarrow$ 1$\times$1, 2] & dilation=1 & ReLU & 64$\times$64 \\
3 & DBNeck \textit{/wo R} & 64$\times$64 & [1$\times$1 $\rightarrow$ 3$\times$3 $\rightarrow$ 1$\times$1, 2] & dilation=1 & ReLU & 32$\times$32 \\
4 & DBNeck \textit{/w R} & 32$\times$32 & [1$\times$1 $\rightarrow$ 3$\times$3 $\rightarrow$ 1$\times$1, 1] & dilation=1 & ReLU & 32$\times$32 \\
5 & DBNeck \textit{/wo R} & 32$\times$32 & [1$\times$1 $\rightarrow$ 3$\times$3 $\rightarrow$ 1$\times$1, 2] & dilation=2 & ReLU & 16$\times$16 \\
6 & DBNeck \textit{/w R} & 16$\times$16 & [1$\times$1 $\rightarrow$ 3$\times$3 $\rightarrow$ 1$\times$1, 1] & dilation=1 & ReLU & 16$\times$16 \\
7 & LoGRA & 16$\times$16 & [3$\times$3 $\rightarrow$ 1$\times$1 $\rightarrow$ 1$\times$1 $\rightarrow$ 3$\times$3] & --- & GELU / Linear & 16$\times$16 \\
8 & DBNeck \textit{/w R} & 16$\times$16 & [1$\times$1 $\rightarrow$ 3$\times$3 $\rightarrow$ 1$\times$1, 1] & dilation=4 & ReLU & 16$\times$16 \\
9 & DBNeck \textit{/w R} & 16$\times$16 & [1$\times$1 $\rightarrow$ 3$\times$3 $\rightarrow$ 1$\times$1, 1] & dilation=1 & ReLU & 16$\times$16 \\
10 & DBNeck \textit{/w R} & 16$\times$16 & [1$\times$1 $\rightarrow$ 3$\times$3 $\rightarrow$ 1$\times$1, 1] & dilation=1 & ReLU & 16$\times$16 \\
11 & LoGRA & 16$\times$16 & [3$\times$3 $\rightarrow$ 1$\times$1 $\rightarrow$ 1$\times$1 $\rightarrow$ 3$\times$3] & --- & GELU / Linear & 16$\times$16 \\
12 & Conv2D & 16$\times$16 & (1$\times$1, 1) & padding=0 & ReLU & 16$\times$16 \\
13 & AdaptiveAvgPool2d & 16$\times$16 & --- & --- & --- & 1$\times$1 \\
\hline
\end{tabular}
\end{table*}

\subsection{Ultra-lightweight Network}
The proposed ultra-lightweight neural network incorporating \texttt{LoGRA} is designed to balance computational efficiency optimally and feature extraction capability. Its architectural overview is illustrated in Fig.~\ref{fig:ufedtomatt}, with detailed layer configurations provided in Table~\ref{tab:ultralightweightnetwork}. The network starts with a \(3 \times 3\) convolutional layer (\(\text{stride} = 2\)), which reduces the input resolution from \(256 \times 256\) to \(128 \times 128\) and reducing computational cost while retaining key low-level features. This is followed by a series of dilated bottleneck (DBNeck) and \texttt{LoGRA} layers.

\subsubsection{\textbf{DBNeck}}
The DBNeck is the primary convolutional building block, responsible for local feature extraction, spatial downsampling and maintaining a lightweight structure. DBNeck layers progressively enhance feature representations. As illustrated in Fig.~\ref{fig:ufedtomatt}, two variants of the DBNeck are utilised: one without a residual connection (DBNeck \textit{/wo R}) and another with a residual connection (DBNeck \textit{/w R}). Regardless, each DBNeck adopts a \(1 \times 1 \rightarrow 3 \times 3 \rightarrow 1 \times 1\) convolution configuration. First, a pointwise \(1 \times 1\) convolution expands the input channels to an intermediate dimension. This is followed by a depthwise \(3 \times 3\) dilated convolution, which enlarges the receptive field without increasing parameters or computational cost and can also perform stride-based downsampling. Finally, a pointwise \(1 \times 1\) convolution projects the intermediate features to the output dimension. Moreover, variable dilation rates (\(1, 2\) and \(4\)) extend the receptive field without increasing the number of parameters and enhance the efficiency of multi-scale feature learning.

\subsubsection{\textbf{LoGRA}}
The architecture integrates \texttt{LoGRA} modules with residual fusion at key stages to strengthen interactions between local and global contexts while preserving information flow across layers. It initially processes the input using depthwise and pointwise convolutions to generate locally enriched features, which, together with residual inputs, are transformed into patch sequences and passed through a lightweight transformer that captures long-range dependencies with linear computational complexity. Each \texttt{LoGRA} block comprises a $3 \times 3 \rightarrow 1 \times 1 \rightarrow 1 \times 1 \rightarrow 3 \times 3$ convolutional sequence with Gaussian error linear unit (GELU) or linear activations, facilitating global reasoning via residual connections while maintaining a lightweight design. The transformer output is reshaped into a two-dimensional feature map and fused with the original input through convolutional operations, forming the final representation. This design effectively integrates fine-grained local details with global structural information, enhances feature coherence and strengthens discriminative representation learning while remaining computationally efficient.

The alternating arrangement of DBNeck and \texttt{LoGRA} modules enables progressive refinement of spatial features and improves representational quality across the network. The extensive use of \(1 \times 1\) convolutions throughout the architecture considerably reduces computational cost and parameter count. An adaptive average pooling layer aggregates spatial features into a compact \(1 \times 1\) representation that serves as the input to the final classification or regression stage. The combination of ReLU activations in convolutional layers and DBNeck layers, along with GELU or linear activations in \texttt{LoGRA} layers, facilitates smooth gradient flow and enhances non-linear expressiveness. The proposed architecture strikes an effective balance between efficiency and performance by coupling dilated convolutional feature extraction with global reasoning, which provides a lightweight yet expressive framework for a broad range of vision tasks.

\section{Experiments} \label{sec:experiments}
In this section, we present the results and discussion of extensive experiments conducted to evaluate the effectiveness of the proposed \texttt{U-FedTomAtt} framework for recognising tomato leaf diseases.

\subsection{Implementation and Parameter Settings}
We implemented our proposed framework using the PyTorch\footnote{https://pytorch.org/} open-source DL platform on a high-performance workstation equipped with dual Intel\textsuperscript{\textregistered} Xeon\textsuperscript{\textregistered} 4215R processors and two NVIDIA RTX A6000 GPUs, each with 48 GB of VRAM. The system was further supported by 128 GB of DDR4 memory and an HP Z Turbo Quad Pro 512 GB SSD, along with dual 4 TB SATA HDDs for efficient data handling and storage.

The model input size was set to $256 \times 256$, and training was conducted using the Adam optimiser with an initial learning rate of $0.001$ and a weight decay of $0.01$ to ensure effective convergence and regularisation. A label smoothing factor of $0.1$ was applied to improve generalisation and reduce overconfidence in predictions. The model was trained with a batch size of $64$ for 10 epochs in each training round, and the entire training procedure consisted of $50$ FL rounds. These hyperparameter configurations were empirically determined. Furthermore, for the FL setting, the number of clients, $K$, is set to $5$.

\subsection{Datasets and Evaluation Metrics}
The proposed \texttt{U-FedTomAtt} framework for tomato leaf disease recognition is evaluated using two publicly available benchmark datasets: a tomato subset of the PlantVillage dataset and the SLIF-Tomato disease dataset. The PlantVillage subset comprises $18,160$ images across ten classes, including nine disease and one healthy class, each consisting of a single tomato leaf on a uniform background with notable class imbalance. Similarly, the SLIF-Tomato disease dataset comprises 8,934 images across eight healthy and diseased categories, representing a fully in-field collection. For experimental evaluation, 20\% of the data is allocated for testing in both datasets. To better emulate real-world federated learning scenarios, the data is deliberately distributed unevenly among clients to enhance its non-IID properties.

The results are reported using two standard performance metrics: \textbf{Top-1} accuracy, \textbf{F1-score}, \textbf{precision} and \textbf{recall}. The Top-1 accuracy measures the proportion of correctly predicted samples, while the F1-score represents the harmonic mean of precision and recall.

\subsection{Results and Discussion}
To demonstrate the efficiency of the proposed ultra-lightweight neural network with the novel \texttt{LoGRA} module, we conduct a series of comprehensive experiments and compare its performance against several lightweight convolutional and transformer-based baselines widely used in vision tasks.

\begin{enumerate}
    \item \textbf{MobileViT-v1} \cite{mehta2021mobilevit} integrates convolutional inductive biases within a lightweight transformer backbone to achieve efficient global–local feature representation while maintaining computational efficiency.
    \item \textbf{DeiT-tiny} \cite{touvron2021training} represents a compact variant of the DeiT family networks, which employs strong data augmentation and distillation strategies to train transformers effectively with limited data.
    \item \textbf{PiT-tiny} \cite{heo2021rethinking} introduces spatial reduction through pooling layers between transformer stages to balance computational cost and representational richness.
    \item \textbf{PSLT-tiny} \cite{wu2023pslt} is a lightweight transformer architecture designed to improve spatial–semantic learning trade-offs.
    \item \textbf{MobileNetV2} \cite{sandler2018mobilenetv2} is a highly efficient CNN-based model that utilises inverted residuals and linear bottlenecks to achieve strong performance with minimal computation.
    \item \textbf{SqueezeNet1\_0} \cite{iandola2016squeezenet} achieves AlexNet-level accuracy with significantly fewer parameters through fire modules and aggressive parameter reduction.
    \item \textbf{ShuffleNetV2} \cite{ma2018shufflenet} enhances computational efficiency through channel shuffle operations that promote effective feature reuse.
\end{enumerate}

Furthermore, to evaluate the proposed FedDAWA aggregation algorithm under FL conditions, we compare it against several standard FL baselines and a centralised non-FL approach.

\begin{enumerate}
    \item \textbf{Non-FL} is centralised training where all data are stored on a single server. It provides the upper bound for performance without privacy or communication limits.
    \item \textbf{FedAvg} \cite{mcmahan2017communication} is the basic FL aggregation technique that averages model parameters after local training. It is simple but often fails under data heterogeneity.
    \item \textbf{FedProx} \cite{li2020federated} adds a proximal term to stabilise local optimisation. It performs better than FedAvg under non-IID data.
    \item \textbf{FedNova} \cite{wang2020tackling} normalises local updates before global aggregation. It improves convergence when clients perform different numbers of local steps.
    \item \textbf{FeDABoost} \cite{arachchige2025fedaboost} uses a boosting principle that reweights client updates based on performance contribution. It enhances robustness against client variation.
    \item \textbf{Floco} \cite{grinwald2024federated} applies a local correction step that aligns updates before aggregation. It reduces client drift and improves accuracy under non-IID data.
\end{enumerate}

\begin{figure}[ht]
    \centering
    \includegraphics[width=\linewidth]{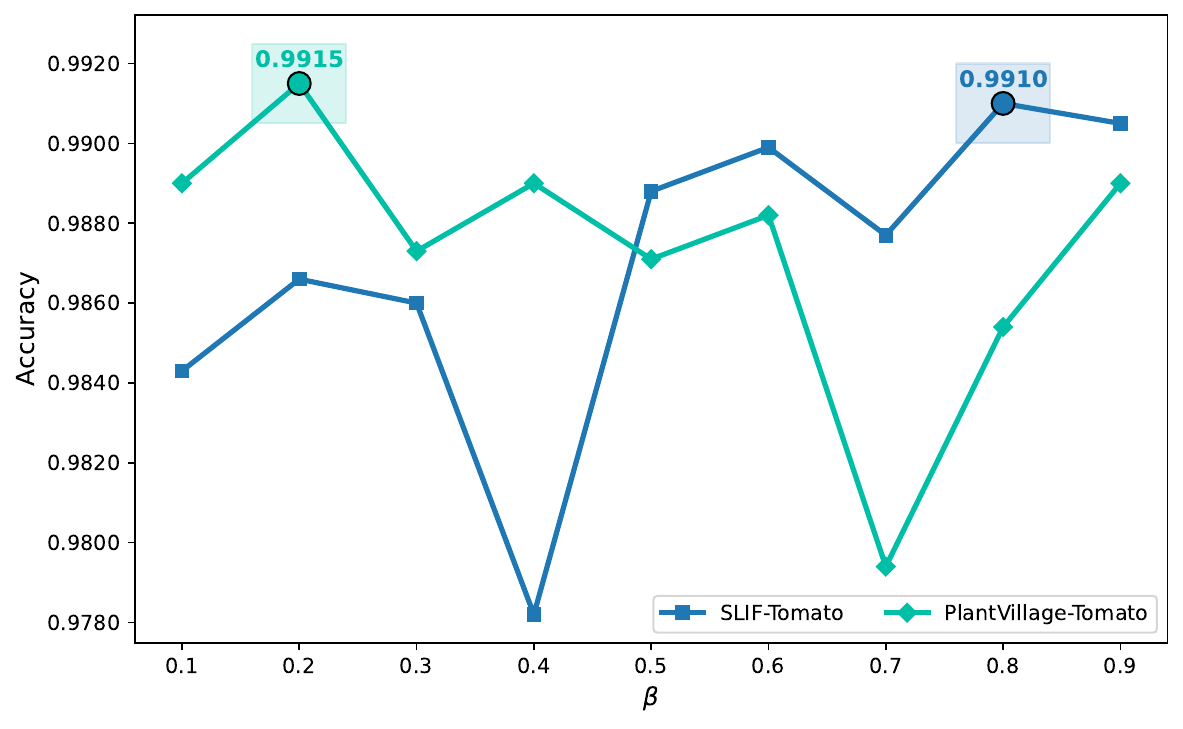}
    \caption{The selection of the hyperparameter $\beta$ for SLIF-Tomato and PlantVillage tomato datasets.}
    \label{fig:beta}
\end{figure}

The hyperparameter \( \beta \), which regulates the balance between dataset size (\( P_k \)) and model performance (\( Q_k \)) in the adaptive aggregation formulation, was empirically optimised through a grid search within the range \( [0.1, 0.9] \). As illustrated in Fig.~\ref{fig:beta}, the performance trends on both SLIF-Tomato and PlantVillage-Tomato datasets exhibit a smooth yet distinct sensitivity to variations in $\beta$. For the SLIF-Tomato dataset, accuracy improves with increasing $\beta$, except at 0.3 and 0.4, reflecting the advantage of incorporating dataset size into the aggregation process. The large $\beta$ values, such as 0.9, diminish the emphasis on model quality and result in a slight decline in accuracy beyond the optimal point. In contrast, accuracy for the PlantVillage-Tomato dataset decreases as $\beta$ increases. The maximum global accuracy for the SLIF-Tomato dataset (0.9910) is achieved at \( \beta = 0.8 \), while the highest accuracy for the PlantVillage tomato dataset (0.9915) is recorded at \( \beta = 0.2 \). These results suggest that the optimal trade-off between data quantity and learning quality depends on the degree of data heterogeneity among clients.

\begin{table*}[htbp]
\centering
\caption{Federated performance comparison (Top-1 Accuracy and F1-score) between the proposed approach and existing state-of-the-art deep networks. The proposed \texttt{FedDAWA} FL aggregation is applied for all networks compared.}
\label{tab:federatedperformance}
\renewcommand{\arraystretch}{1.3}
\setlength{\tabcolsep}{4pt}
\begin{tabular}{|c|
                l|
                P{0.85cm}|P{0.85cm}|
                P{0.85cm}|P{0.85cm}|
                P{0.85cm}|P{0.85cm}|
                P{0.85cm}|P{0.85cm}|
                P{0.85cm}|P{0.85cm}||
                P{0.85cm}|P{0.85cm}|}
\hline
\multirow{3}{*}{\textbf{}} &
\multirow{2}{*}{\textbf{Model}} & 
\multicolumn{2}{c|}{\textbf{Local 1}} & 
\multicolumn{2}{c|}{\textbf{Local 2}} & 
\multicolumn{2}{c|}{\textbf{Local 3}} & 
\multicolumn{2}{c|}{\textbf{Local 4}} & 
\multicolumn{2}{c||}{\textbf{Local 5}} & 
\multicolumn{2}{c|}{\textbf{FL (Global)}} \\ 
\cline{3-14}
 &  & \textbf{Top-1} & \textbf{F1} 
 & \textbf{Top-1} & \textbf{F1} 
 & \textbf{Top-1} & \textbf{F1} 
 & \textbf{Top-1} & \textbf{F1} 
 & \textbf{Top-1} & \textbf{F1} 
 & \textbf{Top-1} & \textbf{F1} \\ 
\hline

\multirow{8}{*}{\rotatebox[origin=c]{90}{\textbf{SLIF-Tomato}}}
& MobileNetV2 \cite{sandler2018mobilenetv2}         & .9961 & .9966 & .9843 & .9867 & .9412 & .9467 & .9485 & .9563 & .7890 & .8046 & .9731 & .9775 \\ 
& SqueezeNet1\_0 \cite{iandola2016squeezenet}       & .9709 & .9745 & .9569 & .9615 & .9692 & .9728 & .9351 & .9434 & .9754 & .9784 & .9748 & .9776 \\
& ShuffleNetV2 \cite{ma2018shufflenet}              & .9759 & .9797 & .9894 & .9910 & .9737 & .9772 & .9743 & .9782 & .9933 & .9943 & .9771 & .9821 \\ 
& PSLT\_tiny \cite{wu2023pslt}                      & .9457 & .9493 & .9855 & .9866 & .9715 & .9732 & .9838 & .9857 & .8987 & .9035 & .9838 & .9863 \\ 
& PiT-tiny \cite{heo2021rethinking}                 & .9933 & .9940 & .9843 & .9867 & .9944 & .9949 & .9172 & .9285 & .9737 & .9772 & .9799 & .9816 \\ 
& DeiT-tiny \cite{touvron2021training}              & .9687 & .9709 & .9591 & .9615 & .9866 & .9874 & .9810 & .9833 & .9564 & .9592 & .9843 & .9855 \\
& MobileViT-v1 \cite{mehta2021mobilevit}            & .9754 & .9788 & .9871 & .9889 & .9938 & .9945 & .9860 & .9873 & .9508 & .9536 & .9894 & .9910 \\ 
\cline{2-14}
& \texttt{U-FedTomAtt} & .9603 & .9643 & .9782 & .9821 & .9821 & .9841 & .9748 & .9789 & .9866 & .9886 & .9910 & .9923 \\ \hline

\multirow{8}{*}{\rotatebox[origin=c]{90}{\textbf{PlantVillage Tomato}}}
& MobileNetV2 \cite{sandler2018mobilenetv2}         & .9948 & .9928 & .9920 & .9985 & .9879 & .9841 & .9782 & .9721 & .9763 & .9674 & .9909 & .9887 \\ 
& SqueezeNet1\_0 \cite{iandola2016squeezenet}       & .9895 & .9862 & .9727 & .9647 & .9755 & .9667 & .9942 & .9917 & .9802 & .9736 & .9909 & .9865 \\
& ShuffleNetV2 \cite{ma2018shufflenet}              & .9816 & .9745 & .9854 & .9804 & .9813 & .9722 & .9766 & .9690 & .9871 & .9826 & .9912 & .9880 \\
& PSLT\_tiny \cite{wu2023pslt}                      & .9788 & .9709 & .9865 & .9817 & .9882 & .9824 & .9846 & .9784 & .9678 & .9519 & .9871 & .9845 \\ 
& PiT-tiny \cite{heo2021rethinking}                 & .9799 & .9748 & .9449 & .9323 & .9681 & .9590 & .9672 & .9583 & .9901 & .9885 & .9846 & .9814 \\ 
& DeiT-tiny \cite{touvron2021training}              & .9584 & .9490 & .9620 & .9537 & .9736 & .9691 & .9648 & .9613 & .9504 & .9417 & .9884 & .9862 \\ 
& MobileViT-v1 \cite{mehta2021mobilevit}            & .9683 & .9606 & .9868 & .9851 & .9835 & .9781 & .9879 & .9856 & .9865 & .9847 & .9912 & .9898 \\ 
\cline{2-14}
& \texttt{U-FedTomAtt} & .9686 & .9623 & .9769 & .9706 & .9851 & .9827 & .9827 & .9781 & .9774 & .9686 & .9915 & .9897 \\ \hline

\end{tabular}
\end{table*}

In Table \ref{tab:federatedperformance}, the Top-1 accuracy and F1-scores are reported for each local client as well as the global federated model to enable a comprehensive assessment of performance consistency across diverse data distributions. All results presented in this table are derived within the FL framework with the corresponding $\beta$ values. For a fair comparison, the proposed \texttt{FedDAWA} is applied to all deep networks compared. The local metrics indicate the model’s performance on each client’s dataset and reflect the global model’s ability to generalise across local variations. For the SLIF-Tomato dataset, the proposed \texttt{U-FedTomAtt} achieves the highest global accuracy, recording a Top-1 accuracy of 0.9910 and an F1-score of 0.9923, outperforming all baseline models. While several models show strong performance locally, \texttt{U-FedTomAtt} maintains the best balance across all clients, which demonstrates superior stability and generalisation under non-IID data. Even with fluctuations in local accuracies across clients, the global performance remains consistently high and confirms the robustness of the proposed architecture in federated aggregation.

Similarly, for the PlantVillage Tomato dataset, \texttt{U-FedTomAtt} once again records the best global performance, with a Top-1 accuracy of 0.9915 and an F1-score of 0.9897. \texttt{U-FedTomAtt} exhibits consistent improvement across local clients and demonstrates strong capability in handling well-curated and balanced datasets effectively. MobileViT-v1 and ShuffleNetV2 achieve competitive local accuracies yet exhibit slightly lower global performance than \texttt{U-FedTomAtt}, which indicates superior collaborative learning efficiency of the latter. The results demonstrate that \texttt{U-FedTomAtt} achieves both local consistency and global superiority within the FL framework. Its stable performance across heterogeneous clients indicates strong suitability for real-world agricultural applications characterised by variable data distributions across devices and locations.

In addition to its superior performance, the proposed \texttt{U-FedTomAtt} exhibits remarkable computational efficiency compared to existing state-of-the-art architectures, as summarised in Table~\ref{tab:efficiency}. With only 71.41 MFLOPs and 245.34K parameters, \texttt{U-FedTomAtt} achieves a substantial reduction in computational and memory requirements relative to other models. For instance, its FLOPs are approximately six times lower than those of MobileNetV2 and ShuffleNetV2, while the parameter count is nearly an order of magnitude smaller than that of MobileViT-v1 and DeiT-tiny. Ultimately, this reduction directly enhances both training and inference efficiency within the FL framework. It facilitates deployment on resource-constrained edge devices such as agricultural sensors and mobile platforms. The model maintains a strong representational capacity despite its lightweight design, achieving competitive Top-1 accuracy and F1-scores across the evaluated tomato disease datasets. The demonstrated balance between computational efficiency and predictive performance affirms the scalability and practical applicability of \texttt{U-FedTomAtt} in real-world agricultural environments characterised by limited computational resources and high reliability requirements.

\begin{figure*}[h]
    \centering
    \includegraphics[width=\linewidth]{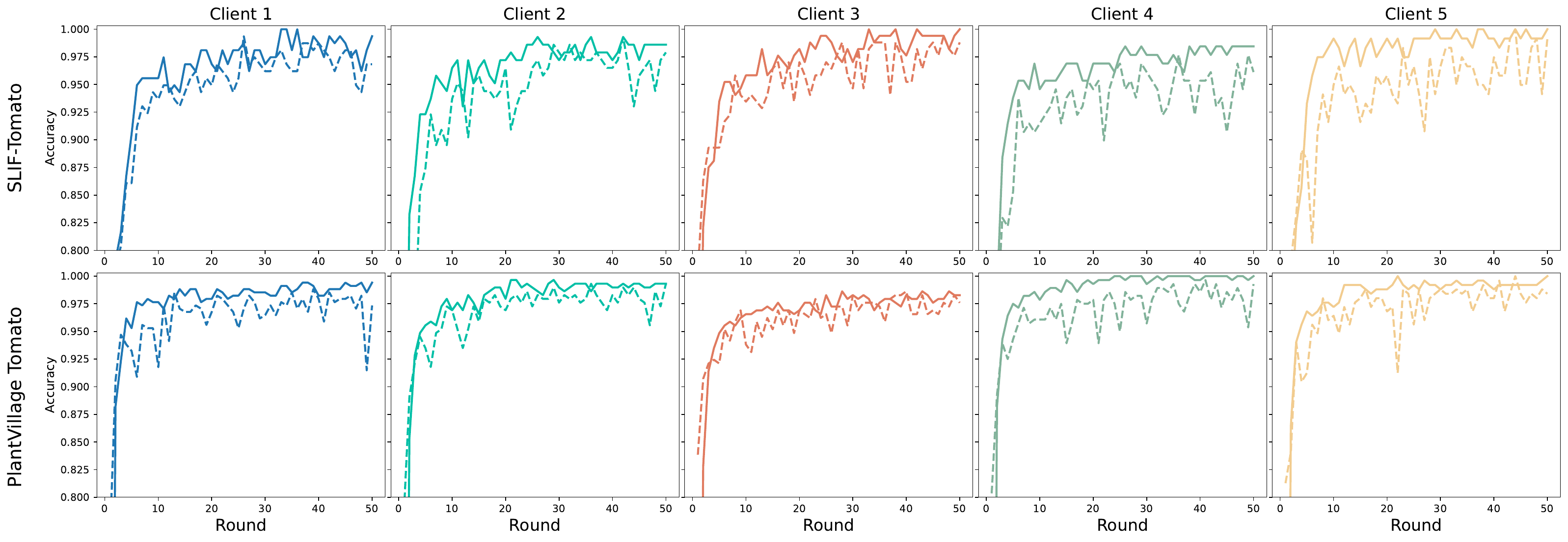}
    \caption{Evolution of local and global model accuracies over 50 communication rounds. The continuous lines and broken lines indicate the accuracy of the global model and local models, respectively.}\label{fig:localglobalaccuracy}
\end{figure*}

Beyond the quantitative results, Figure \ref{fig:localglobalaccuracy} illustrates the evolution of both local and global model accuracies on local validation data of each client over 50 communication rounds. The global model achieves higher validation accuracy than the individual local models, demonstrating its ability to effectively aggregate knowledge across all clients. The variation observed in the accuracy trajectories of different clients reflects the non-IID nature of their data distributions. Local models exhibit fluctuating accuracies due to limited and diverse data, whereas the global model exhibits a smoother, more stable convergence pattern. These results indicate that the federated aggregation process effectively enhances generalisation performance across heterogeneous clients.

\begin{table}[h]
\centering
\caption{Comparison of computational efficiency of the proposed ultra-lightweight neural network against existing state-of-the-art neural networks.}
\label{tab:efficiency}
\renewcommand{\arraystretch}{1.3} 
\begin{tabular}{|l|c|c|}
\hline
\multirow{2}{*}{\textbf{Model}} 
& \multicolumn{2}{c|}{\textbf{Efficiency Metrics}} \\
\cline{2-3}
& \textbf{FLOPs (M)} & \textbf{Params (K)} \\
\hline
MobileNetV2 \cite{sandler2018mobilenetv2}         & 399.95 & 2,234.12 \\
SqueezeNet1\_0 \cite{iandola2016squeezenet}       & 973.30 & 739.53 \\
ShuffleNetV2 \cite{ma2018shufflenet}              & 390.14 & 2,486.82 \\
PSLT\_tiny \cite{wu2023pslt}                      & 874.40 & 3,917.07 \\
PiT-tiny \cite{heo2021rethinking}                 & 652.70 & 4,607.18 \\
DeiT-tiny \cite{touvron2021training}              & 1,406.30 & 5,537.48  \\
MobileViT-v1 \cite{mehta2021mobilevit}            & 449.24 & 1,059.20 \\ \hline
\texttt{U-FedTomAtt}                              & 71.41 & 245.34 \\
\hline
\end{tabular}
\end{table}

\begin{table*}[htbp]
\centering
\caption{Performance comparison of federated aggregation methods and the baseline non-FL approach against the proposed \texttt{FedDAWA}, using the proposed ultra-lightweight neural network.}
\label{tab:federated}
\setlength{\tabcolsep}{3pt}
\renewcommand{\arraystretch}{1.3}
\begin{tabular}{|l|
                >{\centering\arraybackslash}p{1.3cm}|
                >{\centering\arraybackslash}p{1.3cm}|
                >{\centering\arraybackslash}p{1.3cm}|
                >{\centering\arraybackslash}p{1.3cm}||
                >{\centering\arraybackslash}p{1.3cm}|
                >{\centering\arraybackslash}p{1.3cm}|
                >{\centering\arraybackslash}p{1.3cm}|
                >{\centering\arraybackslash}p{1.3cm}|}
\hline
\multirow{2}{*}{\textbf{Method}} & \multicolumn{4}{c||}{\textbf{SLIF-Tomato}} & \multicolumn{4}{c|}{\textbf{PlantVillage Tomato}} \\ \cline{2-9}
 & \textbf{Top-1} & \textbf{F1} & \textbf{Precision} & \textbf{Recall} & \textbf{Top-1} & \textbf{F1} & \textbf{Precision} & \textbf{Recall} \\ 
\hline
Non-FL                                      & .9407 & .9242 & .9242 & .9201 & .9127 & .9208 & .9205 & .9255 \\ \hline
+ FedAvg \cite{mcmahan2017communication}    & .9873 & .9843 & .9851 & .9837 & .9821 & .9853 & .9845 & .9869 \\ 
+ FedProx \cite{li2020federated}            & .9879 & .9860 & .9881 & .9841 & .9849 & .9871 & .9858 & .9889 \\ 
+ FedNova \cite{wang2020tackling}           & .9865 & .9834 & .9865 & .9807 & .9855 & .9881 & .9871 & .9896 \\
+ FeDABoost \cite{arachchige2025fedaboost}  & .9851 & .9808 & .9826 & .9792 & .9815 & .9846 & .9837 & .9862 \\
+ Floco \cite{grinwald2024federated}        & .9849 & .9810 & .9820 & .9801 & .9855 & .9876 & .9866 & .9888 \\
\hline
+ \textbf{\texttt{FedDAWA}} & \textbf{.9915} & \textbf{.9897} & \textbf{.9908} & \textbf{.9887} & \textbf{.9910} & \textbf{.9923} & \textbf{.9914} & \textbf{.9934} \\ 
\hline
\end{tabular}
\end{table*}

The effectiveness of the proposed \texttt{FedDAWA} aggregation algorithm is further validated through comparison with several widely adopted FL aggregation methods, as presented in Table \ref{tab:federated}. Traditional algorithms such as FedAvg~\cite{mcmahan2017communication}, FedProx \cite{li2020federated} and FedNova \cite{wang2020tackling} demonstrate competitive performance on both the SLIF-Tomato and PlantVillage Tomato datasets. However, these methods exhibit limited capability in addressing client heterogeneity and non-IID data distributions, which often lead to suboptimal global convergence. The proposed \texttt{FedDAWA} achieves the highest Top-1 accuracy, F1-score, precision and recall across both datasets, surpassing the next best-performing aggregation methods by a clear margin. On the SLIF-Tomato dataset, \texttt{FedDAWA} records a Top-1 accuracy of 0.9915 and an F1-score of 0.9897, outperforming the second-best FedProx by approximately 0.36\% and 0.38\%, respectively. Similarly, on the PlantVillage Tomato dataset, it achieves a Top-1 accuracy of 0.9910 and an F1-score of 0.9923, demonstrating consistent improvements in all evaluated metrics.

The superior performance of \texttt{FedDAWA} can be attributed to its adaptive weighting mechanism, which dynamically integrates both dataset size and model quality during aggregation. This mechanism is particularly suited to agricultural data, which are inherently heterogeneous due to variations in crop types, imaging conditions and regional factors. By accounting for these disparities, the approach ensures equitable contribution from all clients and reduces the dominance of large and noisy datasets. The dual-weighted aggregation strategy effectively mitigates bias arising from imbalanced data distributions and inconsistent data quality. It further promotes consistent and stable global model updates across heterogeneous agricultural clients. Conventional aggregation methods depend primarily on uniform or size-based averaging, which restricts their effectiveness under highly non-IID and domain-variable conditions typical of agricultural datasets. The consistent performance gains achieved by \texttt{FedDAWA} across datasets confirm its robustness and adaptability in federated learning environments characterised by diverse and region-specific agricultural data.

\section{Conclusion} \label{sec:conclusion}
In this paper, an ultra-lightweight FL framework, \texttt{U-FedTomAtt}, is proposed for privacy-preserving tomato disease recognition in distributed and resource-constrained agricultural environments. By integrating the \texttt{LoGRA} module, the framework effectively enhances feature representation through selective focus on informative regions, which mitigates the accuracy loss typically caused by lightweight architectures. Furthermore, a novel \texttt{FedDAWA} aggregation algorithm is introduced to balance client contributions by jointly considering dataset size and local model performance, resulting in more stable and accurate global model convergence under non-IID conditions. Extensive experiments conducted on two benchmark tomato disease datasets validate that the proposed framework achieves superior Top-1 accuracy and F1-scores compared to state-of-the-art federated and lightweight learning methods. The results further confirm the significance of both \texttt{LoGRA} and \texttt{FedDAWA} in improving model generalisation and communication efficiency. A potential direction for future research involves extending \texttt{U-FedTomAtt} to support multi-crop disease recognition and incorporating temporal crop health data for dynamic decision-making in precision agriculture.

\ifCLASSOPTIONcaptionsoff
  \newpage
\fi

\bibliographystyle{IEEEtran}
\bibliography{refs}

\begin{IEEEbiography}[{\includegraphics[width=1in,height=1.25in,clip,keepaspectratio]{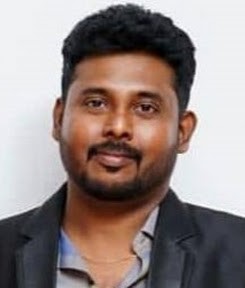}}]{Romiyal George}
 received M.Sc. degree from the University of Peradeniya and a B.Sc. degree from the University of Jaffna, Sri Lanka. He is currently a Ph.D. student at the University of Peradeniya, Sri Lanka.  He has worked in both industry and academia, and his research focuses on machine learning, deep learning, computer vision, and their applications in agriculture, emotion recognition, and steganography.
\end{IEEEbiography}

\begin{IEEEbiography}[{\includegraphics[width=1in,height=1.25in,clip,keepaspectratio]{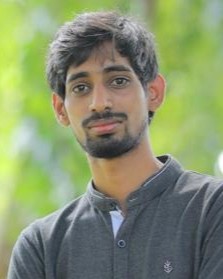}}]{Sathiyamohan Nishankar}
 received the BSc degree in Computer Engineering from the University of Peradeniya, Sri Lanka, in 2023. He is currently a Lecturer with the Faculty of Computing, Sabaragamuwa University of Sri Lanka. His research interests include machine learning, deep learning and computer vision, with applications including 3-dimensional image processing, medical imaging and explainable artificial intelligence.
\end{IEEEbiography}

\begin{IEEEbiography}[{\includegraphics[width=1in,height=1.25in,clip,keepaspectratio]{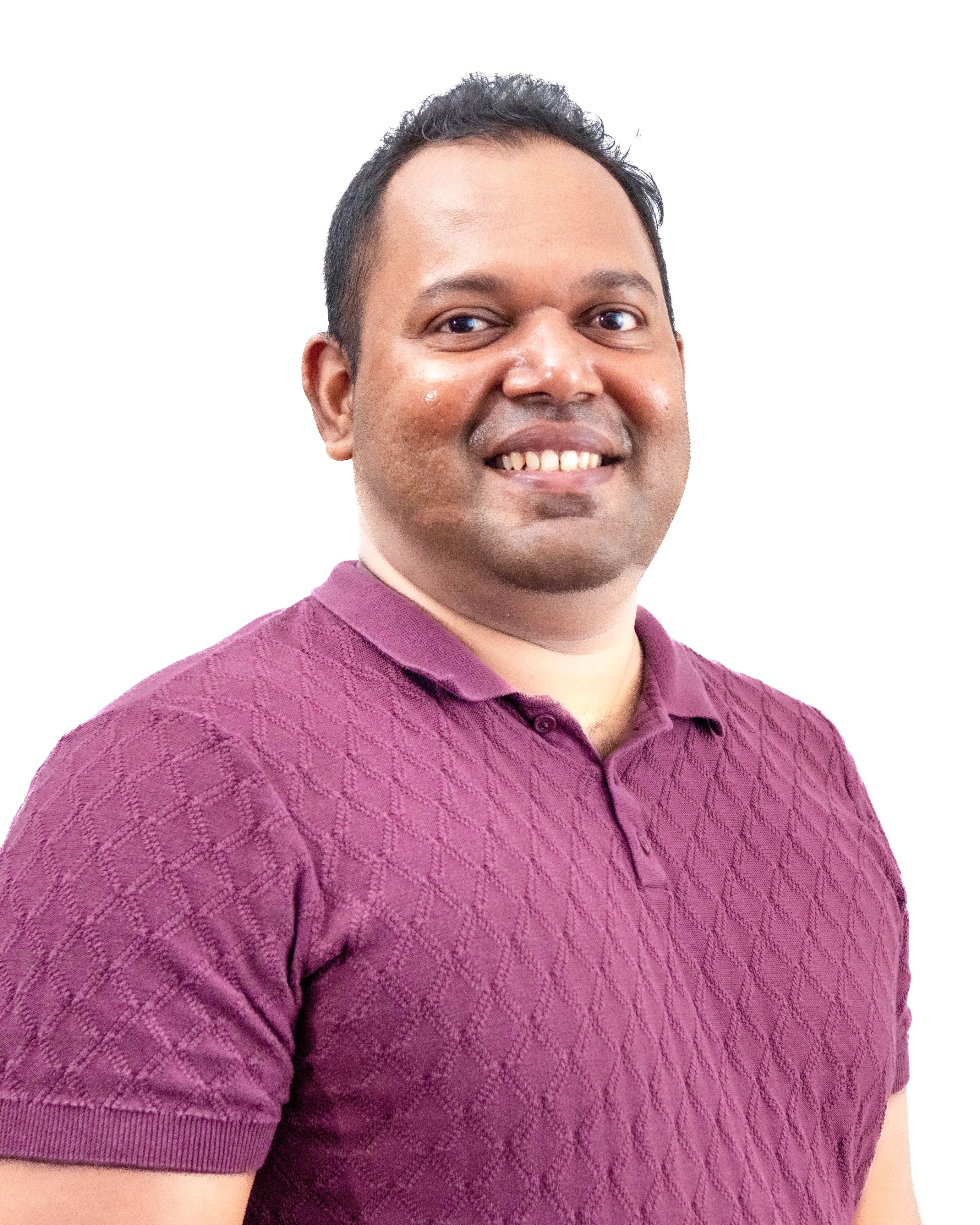}}]{Selvarajah Thuseethan}
	received the Ph.D. degree in information technology from Deakin University, Geelong, VIC, Australia, in 2022. He is currently a Lecturer with the Faculty of Science and Technology, Charles Darwin University, Casuarina, NT, Australia. He was previously a Postdoctoral Research Fellow with the School of Information Technology, Deakin University. His research interests include machine learning, deep learning, computer vision and their applications.
\end{IEEEbiography}

\begin{IEEEbiography}[{\includegraphics[width=1in,height=1.25in,clip,keepaspectratio]{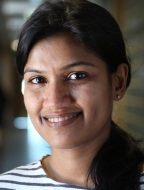}}]{Chathrie Wimalasooriya}
	received her Ph.D. degree from the University of Otago, New Zealand and a Master degree in Software Engineering from the University of Canterbury, New Zealand. She is currently a postdoctoral research fellow with the Canadian Institute for Cybersecurity, University of New Brunswick, Saint John, NB, Canada. Her research interests include repository mining, empirical software engineering, software maintenance, and the application of machine learning. 
\end{IEEEbiography}

\begin{IEEEbiography}[{\includegraphics[width=1in,height=1.25in,clip,keepaspectratio]{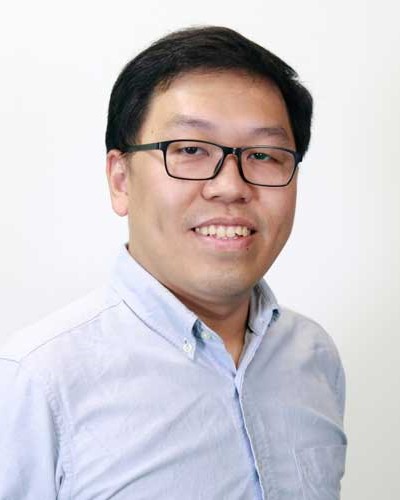}}]{Yakub Sebastian}
	received a Ph.D. degree in Computer Science from Monash University, Clayton, VIC, Australia, in 2016. He is currently a Lecturer in Information Technology with the Faculty of Science and Technology, Charles Darwin University, Casuarina, NT, Australia. His research interests include data science, machine learning, and information retrieval, with applications including precision agriculture and medical imaging.
\end{IEEEbiography}

\begin{IEEEbiography}[{\includegraphics[width=1in,height=1.25in,clip,keepaspectratio]{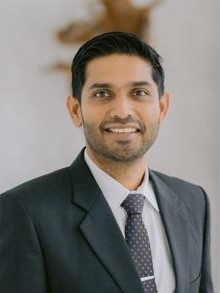}}]{Roshan G. Ragel}
	received his Ph.D. in Computer Science and engineering from UNSW Sydney, Sydney, NSW, Australia. He is a Professor with the Department of Computer Engineering, University of Peradeniya, Sri Lanka. His current research interests include systems-on-chip, the Internet of Things, accelerated and high-performance computing, computational biology, and wearable computing. He has been a Professional Member of the IEEE and IEEE Computer Society since 2005 and a Senior Member since 2014.
\end{IEEEbiography}

\begin{IEEEbiography}[{\includegraphics[width=1in,height=1.25in,clip,keepaspectratio]{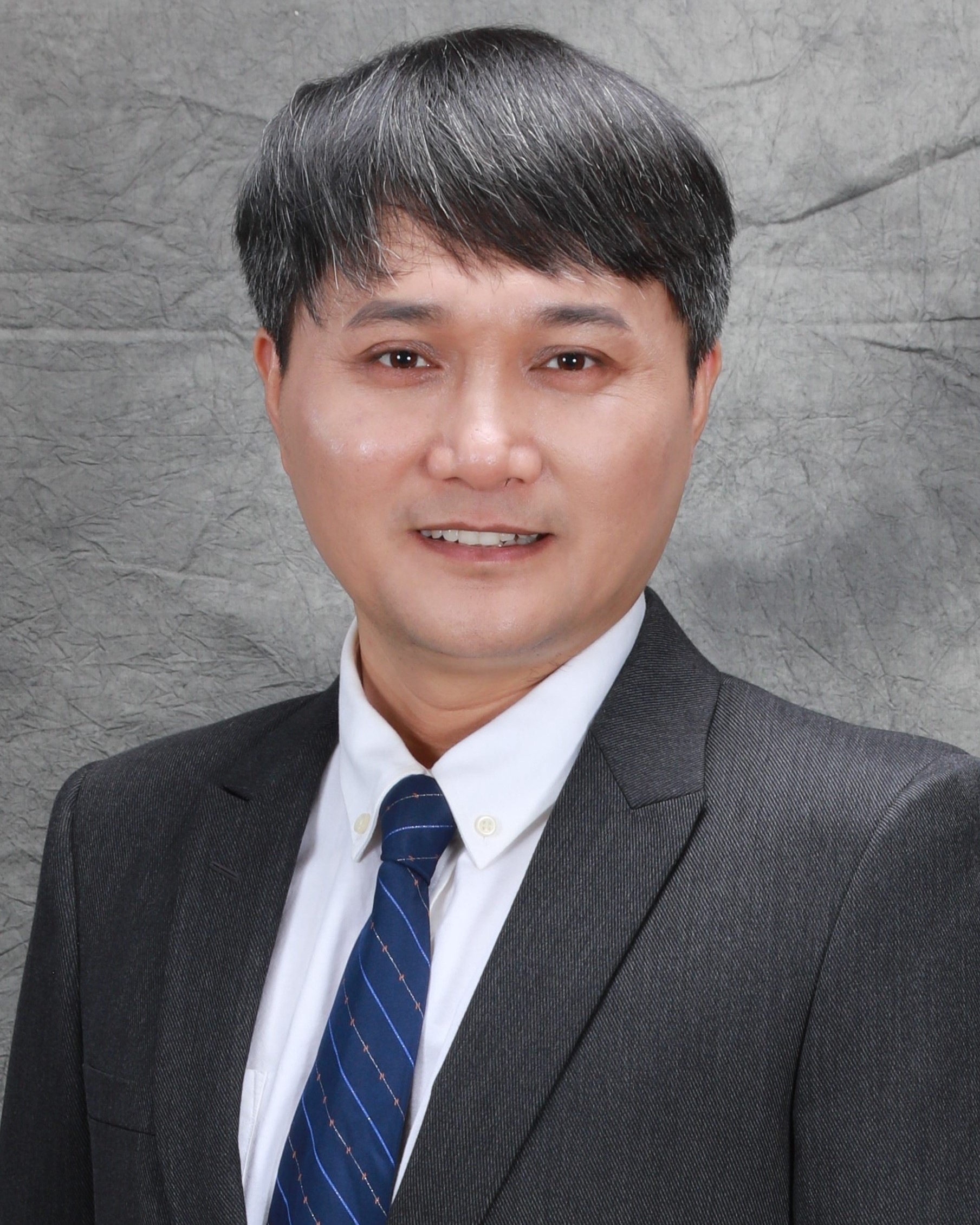}}]{Zhongwei Liang}
	received the Ph.D. degree in Machinery Manufacturing and Automation from the South China University of Technology, China, in 2008, where he also completed postdoctoral research in 2014. He is a Professor and Vice Dean at Guangzhou University, China. He was previously a Research Fellow and Visiting Research Professor with Zhejiang University, China and the University of New South Wales, Australia. His research interests include manufacturing mechanics, fluid mechanics, and advanced manufacturing technologies.
\end{IEEEbiography}

\end{document}